\documentstyle[aps,prb,epsfig,multicol]{revtex}

\renewcommand{\vec}[1]{{\bf #1}}

\newcommand{\wm}{\omega_{\rm m}}
\newcommand{\vn}{\nu_{\rm n}}

\newcommand{\nought}{{\rm o}}

\newcommand{\be}{\begin{equation}}
\newcommand{\ee}{\end{equation}}
\newcommand{\bea}{\begin{eqnarray}}
\newcommand{\eea}{\end{eqnarray}}

\newcommand{\ea}{\mbox{\em et al.}}

\def\ybcoAB{Y$_2$Ba$_4$Cu$_7$O$_{15}$}
\def\ybcoA{YBa$_2$Cu$_3$O$_{7}$} 
\def\ybcoB{YBa$_2$Cu$_4$O$_{8}$}

\newcommand{\rfig}[1]{Fig.~\ref{#1}}
\newcommand{\req}[1]{Eq.~\ref{#1}}

\newcommand{\cut}[1]{}

\begin{document}
\draft
\title{%
Magnetic fluctuations in coupled inequivalent Hubbard layers as a
model for \ybcoAB
}
\author{G.~Hildebrand$^1$, E.~Arrigoni$^1$, J.\ Schmalian$^2$, and W. Hanke$^1$}
\address{$^1$Institut f\"ur Theoretische Physik, 
Universit\"at W\"urzburg, Am
Hubland, D-97074 W\"urzburg, Germany}

\address{%
$^2$Department of Physics, University of Illinois at
Urbana-Champaign, 1110 W. Green Str., Urbana 61801, IL}

\maketitle
\begin{abstract}

We investigate, within the fluctuation-exchange approximation,
a correlated-electron model for \ybcoAB\ represented 
by  two  inequivalent Hubbard layers coupled by an interlayer
hopping $t_\perp$.  An energy offset
$\delta$  is introduced in order to produce a
different charge carrier concentration in the two layers. 
We compare several single-particle and magnetic excitations, namely,
the single particle scattering rate, the spectral function
and the spin lattice as well as spin-spin   relaxation times
in the two layers as a function of $\delta$.
We show that the  induced interlayer magnetic coupling  
produces a tendency to ``equalization'' of the magnetic properties in the
two layers whereby 
antiferromagnetic fluctuations are suppressed in 
the less doped layer and enhanced in the
heavily doped one.
The strong antiferromagnetic bilayer coupling  
causes the charge carriers in the plane
  with larger doping concentration to behave similar to those of the underdoped 
 layer, they are  coupled to. This effect grows for decreasing temperature.
 For high temperatures or if both layers are optimally or overdoped,
i.e.  when the antiferromagnetic correlation length becomes of the
order or smaller than one
  lattice site   the  charge carrier and magnetic dynamics 
of the two layers is disconnected and the equalization effect disappears. 
These results are in good  agreement with
  NMR experiments on \ybcoAB\
by Stern {\em et al.} {\footnotesize 
Phys. Rev B {\bf 51}, 15478  (1995)%
}.
We also compare the results with calculations on bilayer
systems with equivalent layers
as models for the constituent compounds
\ybcoA\ and \ybcoB.

\end{abstract}
\ifpreprintsty\onecolumn\fi

\begin{multicols}{2}
\section{Introduction}
\label{intro}

There is a large amount of  consensus that
the anomalous properties of cuprate superconductors
are caused by the strongly interacting   electrons
within the CuO$_2$ planes.  However, in particular  the observation that
the highest superconducting transition temperatures belong
to   compounds with more than one layer per unit cell
initiated various investigations of out-of-plane properties.
The observation of a rather strong  coupling between 
adjacent layers  has been  made by inelastic neutron
scattering~\cite{tr.ge.92}   (INS), nuclear magnetic 
resonance~\cite{st.ma.94,st.ma.95.1,st.ma.95.2,mo.ri.94} (NMR)
and indirectly also in Raman scattering
experiments~\cite{mo.ch.95}.
Furthermore, the observation of a qualitatively different behavior
of the odd and even channel in INS including a sharp resonance
feature, found solely for odd excitations~\cite{bo.fo.97} and  of a bilayer 
splitting of the Fermi surface  found in  angular resolved photoemission 
experiments (ARPES)~\cite{sc.pa.98.1,sc.pa.98.2}
demonstrate that   low energy excitations 
of   cuprates are affected by the presence of more than one
layer per unit cell.
Related to these issues is the interesting question of the c-axis transport
and the occurrence of a c-axis Josephson plasma excitation
\cite{ande.sci.98,legg.98}, which may
 turn out to be a new probe of    the vortex statics and dynamics of
the superconducting state. 

A very interesting  perspective on the nature of the coupling between 
CuO$_2$-layers  was offered by NMR experiments  by Stern {\em et al.}
on \ybcoAB\ (247). This  material  has a variety of
structural similarities to the extensively studied  
 \ybcoA\ (123) and \ybcoB\ (124) systems.
The main difference  in the crystallographic structure of  123 and 124 
is the double CuO chain in the latter. 
The compound 247 can be considered as a natural multilattice, 
consisting of alternating 124 and 123 blocks.
The bilayers in 247 are correspondingly build up of one
CuO$_2$ layer which belongs to the 123 block and one
 layer of the 124 block.
Based on the analysis of the  NQR  spectra it turned out that the 
charge carrier content in these nonequivalent adjacent layers
is very close to that of the related parent compounds of the two blocks,
i.e. one   plane  has a similar charge carrier concentration 
to the  slightly overdoped 123 system whereas the other layer
corresponds to the underdoped 124 system.
Interestingly, the highest transition temperature (T$_c = 95 \, {\rm K}$)
is  for  247, which has to be compared with the $92 \, {\rm K}$
for 123 and $82 \, {\rm K}$ of the 124 system.
The main experimental  observations of Ref.~\onlinecite{st.ma.94,st.ma.95.1}
are the following: {\rm (i)} the low-temperature  Knight-shift suppression 
 for both  planes is, despite their
different  charge carrier concentration, similar and  behaves like in the 
underdoped  124 system, even though  the high temperature values
of the Knight shift  are the same as
 in the corresponding  123 and 124 compounds. 
{\rm (ii)} both  $^{63}$Cu
 spin lattice  relaxation times of the two  different planar Cu-sites    
 show a spin  pseudogap
in $1/^{63}T_1T$, even though it is barely present in the 123
parent compound itself.
{\rm (iii)} the interplane transverse relaxation rate, 
as measured in a spin-echo
double resonance experiment (which  characterizes the interplane
magnetic susceptibility)  increases for decreasing temperature faster
than the intraplane relaxation rate.
Thus, the main conclusions from these observations are that,
for high temperatures,
the two planes are rather disconnected and behave similarly 
to their parent compounds,
whereas for  decreasing temperatures, the increasing interlayer
magnetic coupling enforces
even the slightly overdoped plane to behave like an underdoped system.

For a proper interpretation of these  interesting experimental data and,
in a more  broader context, for a better understanding 
of the  bilayer coupling in cuprate superconductors in general,
it is essential to investigate to what extent one can describe  
 the main trends
of these data within a model of coupled layers, only different by their
charge carrier concentration or whether one needs to make  qualitatively new 
assumptions about the nature of the bilayer coupling.

One promising approach for the description of bilayer phenomena is 
based on a Hubbard Hamiltonian  with local repulsive Coulomb interaction,
where the interplanar coupling is caused solely by an interplane
hopping element $t_\perp$. This model, restricted to the case of {\it
  equivalent} layers ($\delta=0$), 
has been investigated within various
techniques \cite{bu.sc.92,bu.sc.96,he.ha.94}.
Additional insight  can be gained using 
a self-consistent summation of bubble and ladder diagrams (fluctuation exchange
approximation).
The main results of these
investigations~\cite{da.te.95.2,da.te.96,gr.sc.97.1,gr.sc.97.2}   
are enhanced antiferromagnetic
spin fluctuations due to layer coupling causing,  
in bilayer systems as well,   a d$_{x^2-y^2}$ symmetry of the
superconducting order parameter, a predominantly incoherent
low energy c-axis charge transport even though the bilayer splitting stays
intact, and an enhancement of the relative strength of interlayer vs. intralayer
coupling for decreasing 
doping.

In this paper, we additionally consider the effect of 
an energy offset $\delta$,   which produces a
different charge carrier concentration in the two Hubbard layers
(cf. also  Scalettar et al. \cite{sc.ca.94}).
This is a suitable model to describe the peculiarity of the 247 compound
\ybcoAB, whose bilayers are  build up of one
CuO$_2$ layer  belonging to an \ybcoA\ block and one
 to a \ybcoB\ block.
We evaluate several single-particle and magnetic excitations,
namely, the single-particle
scattering rate, the spectral function
and the spin lattice as well as spin-spin   relaxation times
in the two inequivalent layers
as a function of $\delta$ within the fluctuation-exchange approximation.
We show that the  interlayer coupling  
produces a tendency to equalization of the antiferromagnetic properties in the
two layers whereby 
antiferromagnetic fluctuations are suppressed in 
the less doped layer and enhanced in the
heavily doped one.
 This equalization  effect turns out to be 
enhanced in the presence of 
 antiferromagnetic fluctuations in the system and to be almost absent
 when the antiferromagnetic correlation length becomes of the order or
 smaller than
 one lattice site  and to ultimately decrease for
increasing temperature. 
These results are in good qualitative agreement with
  NMR experiments on \ybcoAB\
by Stern {\em et al.}\cite{st.ma.94,st.ma.95.1}.
We also compare the results with calculations on bilayer
systems with equivalent layers as models for the constituent compounds
\ybcoA\ and \ybcoB.

A first theoretical investigation  of the experimental findings of
Ref.~\cite{st.ma.95.1} has been given by Millis and Monien~\cite{mi.mo.96},
who could  determine the size of the interlayer exchange coupling
from an analysis of the  interlayer cross relaxation time.
These authors also discuss  
that the 41 meV excitation observed in superconducting 
YBa$_{2}$Cu$_{3}$O$_{7}$
is a collective mode pulled down below the superconducting gap by interactions,
and that the observed antisymmetry under interchange of planes follows from the
non-negligible value of $J_{\perp}$. 
An   analysis of the coupling between an undoped layer and an
underdoped one, similar in spirit to ours,  has been carried 
out by Scalettar et al. \cite{sc.ca.94}. 
These authors  study the pairing mechanism, which arises from
the coupling of holes in doped layers to spin fluctuations in the
undoped layers in analogy with the Ginzburg-type scenario for the
coupling of electrons through excitons in a doped semiconductor.
However, it turns out that magnetic fluctuations in the undoped layer
are strongly suppressed by the coupling with the doped layer and 
superconducting correlations are reduced by the interplane coupling at
least at the temperatures accessible to the simulations.
The study of the coupling between a strongly antiferromagnetic  
and a doped  subsystem has some similarities with    the ``stripe scenario''
where hole-poor antiferromagnetic 
regions  are 
considered to be in contact with hole-rich superconducting regions.

The paper is organized as follows:   In Sec.~\ref{mode}
we present
our model for coupled layers with different charge carrier concentration and
summarize the main concept of the fluctuation exchange approximation,
used for the approximate investigation of the model.
In Sec.~\ref{resu} we present our numerical results with particular emphasis
to the single-particle and magnetic fluctuations in the two layers and
focus on the  anisotropy and on the tendency of equalization
of this effects. In order to make contact with the
experimental investigations  on the 247 system, we discuss at length the
temperature dependence of various NMR quantities in Sec.~\ref{rela}.
Finally our results are summarized in Sec.~\ref{conc}.

\section{Model and technique}
\label{mode}

In order to  describe the strong electronic
correlations in the high-$T_c$ superconductors and the
particularities of the system \ybcoAB,
 consisting of  two layers with different charge carrier concentration,
 we use  a system of two 
 two-dimensional Hubbard layers  coupled by an
 hopping element $t_\perp$. 
After   Fourier transformation of the intraplane sites into  
momentum space with in-plane momentum  $\vec k$, 
  the Hamiltonian  reads:
\be
H= \sum_{\stackrel{l_1,l_2}{\vec{k},\sigma}} \left[ H_\nought(\vec k)
\right]_{l_1,l_2}
c^\dagger_{\vec{k}, l_1,\sigma}
c^{}_{\vec{k}, l_2,\sigma}
+ U \sum_i n_{i, \uparrow}  n_{i, \downarrow} \; ,
\label{eq:Hubbard_Hamiltonian}
\ee
where 
$c^\dagger_{\vec{k}, l_1,\sigma}$ creates a particle with spin $\sigma$ at
momentum $\vec k$ in layer $l_1$. Furthermore,
$n_{i, \uparrow}$ is the density operator
at lattice site $\vec R_i$ and spin $\uparrow$, and
$H_\nought(\vec k)$ the Hamilton matrix for the noninteracting system
\be
H_\nought(\vec k) =
\left( \begin{array}{c c}
\epsilon_{\vec k} -\mu& t_\perp\\
t_\perp & \epsilon_{\vec k} + \delta -\mu
\end{array} \right) \; .
\ee
In order to describe theoretically 
a different charge carrier concentration
in the two layers, we additionaly introduce 
an  on-site energy $\delta$  
   in the second layer,  effectively modifying its chemical potential.
The planes are coupled solely through a bare interplane hopping $t_\perp$.
 Furthermore, the bare energy dispersion in each plane is 
\bea
\epsilon_{\vec{k}} &=& -2 t  \left( \cos k_x  + \cos k_y \right)
-4 t'  \cos k_x   \cos k_y\nonumber \\
&&-2 t'' \left[ \cos(2 k_x) + \cos(2 k_y) \right] \; .
\label{eq:bare_dispersion}
\eea
 thus including  second and third-neighbor
 hopping processes $(t',t'')$ to
better model the Fermi surface for the system under consideration.
In  the following calculation, we always set $t_\perp/t=0.4$, 
$U=4t$ \cite{gr.sc.97.1,gr.sc.97.2}
and measure the energies in units
of the next-nearest neighbor hopping $t$.

A diagonalization of $H_\nought(\vec k)$
leads to the bonding and anti-bonding bands
of the noninteracting system
\be
\epsilon^\pm_{\vec k} = \epsilon_{\vec k} + \frac{\delta}{2} \pm
\sqrt{\frac{\delta^2}{4} + t_\perp^2}\;.
\ee
The  single-particle excitations and the thermodynamic properties
 of the interacting system are deduced from the 
Green's function $G(\vec k, i \wm)$ 
obtained through
Dyson's equation which for a two-layer system generalizes to a 
$(2 \times 2)$ matrix
equation
\be
G^{-1}(\vec k, i\wm)=(i\wm+ \mu ) {\bf 1}
-H_\nought (\vec k) \; -\Sigma(\vec k, i \wm) \:.
\label{eq:tdl_dyson}
\ee
Approximations are introduced by 
the explicit choice of the self-energy $\Sigma(\vec k, i \wm)$.
Here, we use the expression for the 
self-energy given by the FLEX approximation \cite{bi.sc.89} 
without 
particle-particle vertex contributions.
Within the FLEX, the irreducible particle-particle vertex is solely
the repulsive Coulomb interaction $U$ and  consequently  irrelevant.
Interference effects between the particle-particle and particle-hole
channel, which may be of relevance for an understanding of the pseudogap
state of underdoped cuprates at low temperatures,
 are beyond the scope of this paper.

Introducing the shorthand notation
$k\equiv (\vec k, i\wm)$ and $q \equiv (\vec q, i\vn)$ for convenience, 
the matrix for the self-energy $\Sigma_{ll'}(k)$ reads
\begin{equation}
\Sigma_{ll'}(k) =
\frac{1}{\beta N}\sum_{k'} V_{ll'}(k - k')  G_{ll'}(k') \, ,
\label{sigma}
\end{equation}
where $\beta=\frac 1{k_B T}$ is the inverse temperature, and
the effective interaction $V_{ll'}$ 
results from an infinite series over spin- and
charge-fluctuations  and is given by
 \begin{eqnarray}
\label{eff_int_ns}
V(q)&=& \frac{3 U^2}{2}
(1-U \chi(q))^{-1}
  \chi(q)  \nonumber \\ & &+
\frac{U^2} {2}(1+U \chi(q))^{-1}
  \chi(q)\\
&&-U^2\chi(q) \nonumber  \;.
\end{eqnarray}
Note, that  $V$ and $\chi$ are $(2\times2)$ matrices, i.e. matrix inversion
and multiplications have to be used.
The bare particle-hole bubble $\chi_{ll'}(q)$ consists
of dressed Green's functions
\begin{equation}
\label{chi}
\chi_{ll'}(q)= -\frac{1}{\beta N}
\sum_{k}G_{ll'}(k+ q)
 G_{l'l}(k) \; .
\end{equation}
In \req{chi}, the Green's functions are determined self-consistently
by solving the set of coupled equations
Eqs. (\ref{eq:tdl_dyson},\ref{sigma},\ref{eff_int_ns},\ref{chi}).
During the self-consistency cycle, we fix the on-site energy $\delta$
and the particle number $n_1 = 1 - x_1$ of the first layer, while the
chemical potential $\mu$ and the particle number of the second plane
$n_2=1-x_2$ are determined at each step. 
 It turns out
that the total particle number $n=n_1+n_2$ 
does not essentially  depend on 
temperature, 
which makes  the physical interpretation of our numerical results 
more straightforward.
To avoid the uncertainties related to a numerical analytical continuation
of correlation functions 
from the imaginary Matsubara to real frequencies, we 
use the recently proposed real-frequency approach to the FLEX
approximation.\cite{sc.la.96.2}

In the following, we shall mainly focus our attention  on two different
parameter sets
in order to mimic a situation with strong and weak antiferromagnetic
fluctuations, respectively.
Specifically, we use a parameter set, for simplicity labeled by ``A'',
with 
$t'=-0.38t, t''=-0.06t$ and
a second one, labeled by ``B'', with
$t'=-0.20t, t'=0.15t$. 
As will be shown below, the parameter set A corresponds to  a system
with pronounced antiferromagnetic fluctuations at low temperature, 
while B has much weaker ones.

\section{Results}
\label{resu}
\begin{figure}
\centerline{\epsfig{file=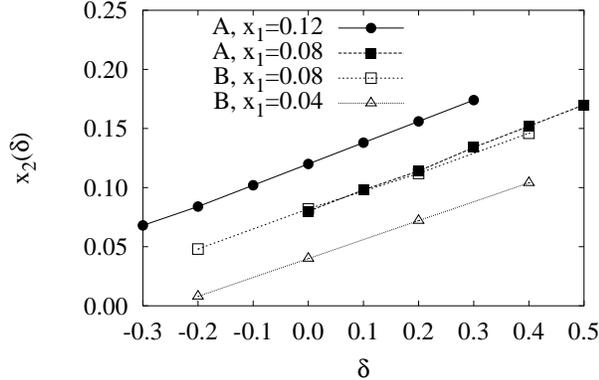,width=8cm}}
\narrowtext
\caption{\label{fig:occupation_vs_ose}
Hole
density
of the second layer as a function of its on-site energy $\delta$ for 
parameter sets A and B (described in the text) 
and different dopings of the first layer $(T=0.02t)$.}
\end{figure}

In order to  investigate the effects of different charge carrier concentrations in coupled
bilayer systems, we start the discussion with the  $\delta$ dependence of 
  the doping of the second plane, $x_2(\delta)$.
Here,  the doping of the first plane $x_1$ and the on-site energy $\delta$
are independent variables for us, while 
$x_2$ comes out from the self-consistent calculation.
 In \rfig{fig:occupation_vs_ose} we present $x_2(\delta)$ for
both parameter sets A and B for different doping levels of the first plane
as a function of $\delta$.
From \rfig{fig:occupation_vs_ose} it can be seen that the
hole doping of the second plane is directly proportional to its on-site
energy $\delta$.
Note that the constant of proportionality,
i.e. the slope $\partial x_2/\partial \delta$, 
 is almost   independent on $t', t''$ and $x_1$,  because the particle
number difference $\delta n = n_1 - n_2$ between both planes
is mainly governed by the energy dependence of the  
effective "chemical potentials" $\mu$ and $\mu-\delta$. On the other
hand, 
correlation
effects do  play  a role here, since the slope 
 depends on $U/t$.
Note that
 this energy difference is determined self-consistently  in our theory 
and therefore does not only depend on $\delta$,
 but varies, due to   self-energy
renormalizations,  also with the strength of the Coulomb interaction $U$.
The reason why 
we follow the
strategy of keeping the doping of the first plane fixed and  only
change
the doping of the second plane is that we want 
to investigate a possible induced 
coupling effect between the planes. 
The question is whether
 single-particle and magnetic properties of the first plane 
are influenced by the doping of the second plane.

\begin{figure}
\centerline{\epsfig{file=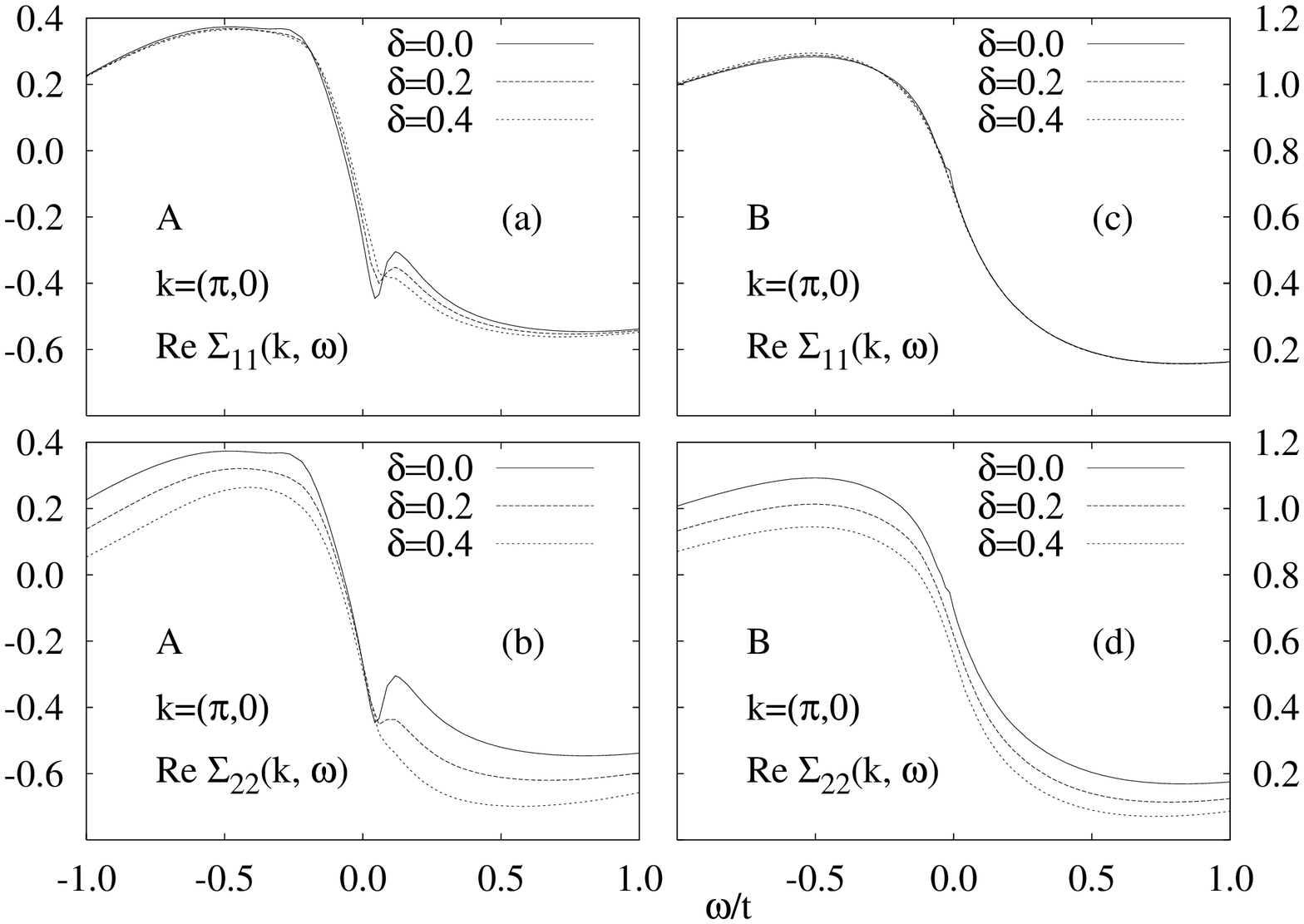,width=9cm}}
\caption{\label{fig:tdl_ReSelfenergy_p0} 
Real part of the diagonal elements
of the self-energy, ${\rm Re}\Sigma_{ll}(\vec k, \omega)$ for $\vec
k=(\pi,0)$ and 
$x_1=0.08,T=0.02t$. (a) and (b) correspond to the parameter set A and
(c) and (d)  to the set B, respectively. }
\end{figure}
In 
\rfig{fig:tdl_ReSelfenergy_p0}, we show the real part of
the diagonal elements of the self-energy $\Sigma_{l,l}(\vec k, \omega)$
with $\vec k=(\pi,0)$ for both layers and both parameter sets A and B.
The different curves in each panel are for various on-site energies
$\delta$. 
\rfig{fig:tdl_ReSelfenergy_p0}(a) displays ${\rm Re}\Sigma$ for the
first (less doped)
layer with parameter set A, i.e. for $t'=-0.38t, t''=0.06t$.
This figure shows that
 ${\rm Re}\Sigma_{11}$ indeed depends  on $\delta$ although the hole
 concentration is not changed by $\delta$ in the first layer.
This demonstrates
that feedback effects due to the interlayer coupling modify also the properties
 of the plane where the
charge carrier concentration is kept constant.
The dip-like structure
for small $\delta$  is a precursor of  new
quasiparticle  states on the shadow of the Fermi surface due to strong
antiferromagnetic fluctuations. This effect, which is strongest for the case
of equivalent planes $(\delta =0)$, has been discussed in Refs.
\onlinecite{gr.sc.97.1,gr.sc.97.2}. An increase of $\delta$ and hence of
the total hole
doping of the system leads to a decrease of the interplanar
antiferromagnetic coupling and thus of the dip structure.
Panel (b) in this figure shows 
${\rm Re}\Sigma_{22}((\pi,0),\omega)$
 for the second
(heavily doped) layer.
Since it is this plane which is primary altered by $\delta$
the influence of changing the on-site energy
$\delta$ is, as expected,  considerably more pronounced. 
These results show that the two layers are strongly connected and
a change in carrier concentration of the second layer strongly
influences the single-particle properties of the first layer as well, 
although
the doping is unchanged here.
However,  whether the two planes are connected or not 
depends on the values of the  parameters of the model.
For example,
a completely different situation is found for 
the parameter set labeled by B.
For this choice, a variation of $\delta$ influences the
second layer [panel (d) of \rfig{fig:tdl_ReSelfenergy_p0}],
but has no effects on the self-energy of the
quasi-particles in the first layer [panel (c)], indicating independent
 planes.

\begin{figure}
\centerline{\epsfig{file=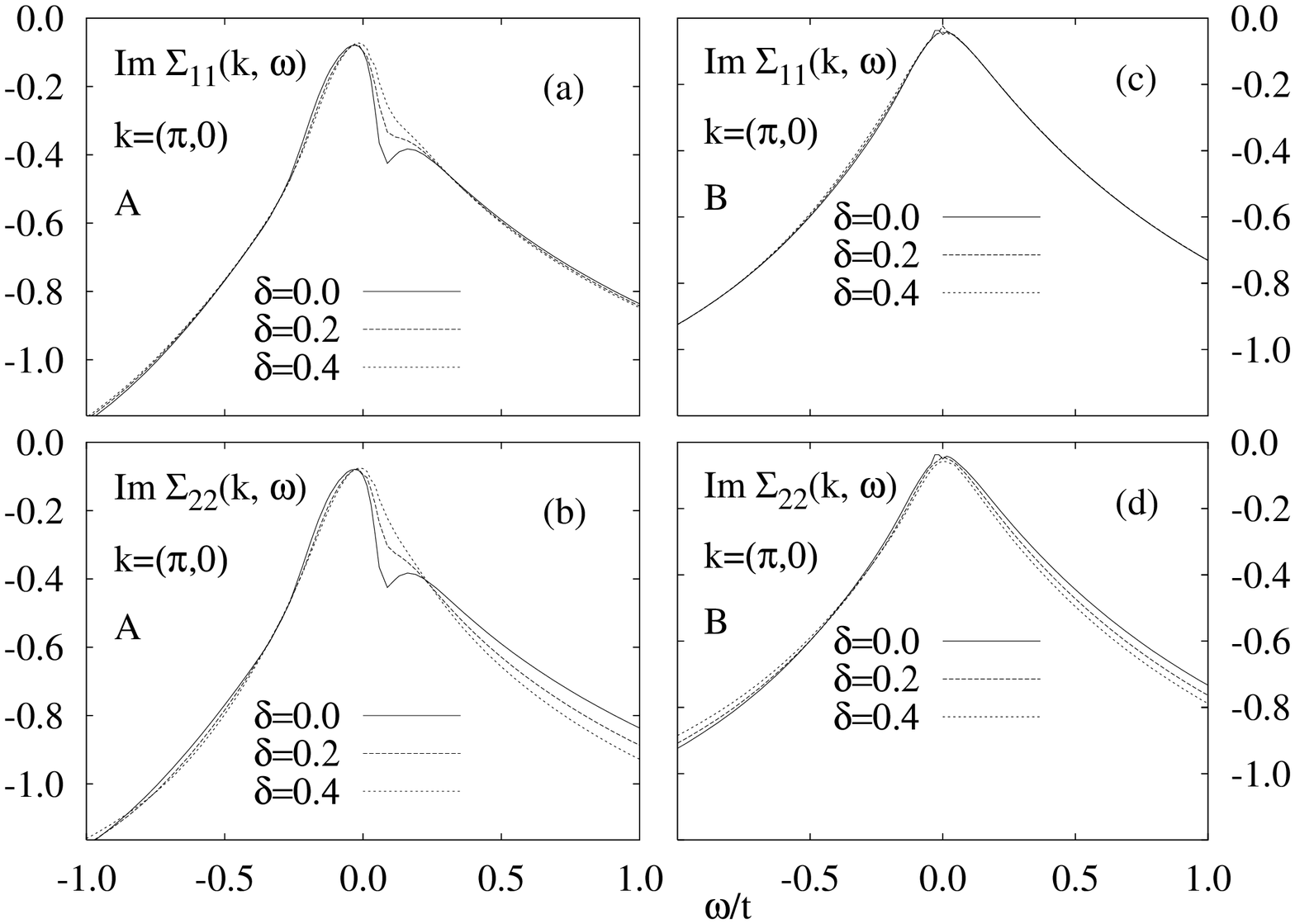,width=9cm}}
\caption{\label{fig:tdl_ImSelfenergy_p0} 
Imaginary part of the diagonal
elements of the self-energy, ${\rm Im}\Sigma_{ll}(\vec k, \omega)$ for
$\vec k=(\pi,0)$ and $x_1=0.08, T=0.02t$}
\end{figure}
The connection between the two planes is also visible in the scattering rates
which are related to the imaginary part of the self-energy and presented 
 for both layers and both parameter
sets in \rfig{fig:tdl_ImSelfenergy_p0}.
For the imaginary part of the self-energy, 
which is more sensitive to low energy excitations, 
we observe an even closer connection between the 
  two
planes for parameter set A
 than for the real parts. 
In addition, these figures demonstrate that precursors of a spin density 
wave state around  $(\pi,0)$,  for low $\delta$,  are rather incoherent due to the
strong scattering rates at these energies. Note also, that even the
quasiparticles at the chemical potential $(\omega=0)$ suffer strong
scattering, as indicated by the rather large values 
${\rm Im}\Sigma_{ll}((\pi,0),\omega=0)$.
Similarly to the real part, the changes in the scattering rates
caused by $\delta$ are rather moderate for parameter set B,
 even in the second layer which is
directly altered by $\delta$ through its doping.
Thus, both the real and the imaginary parts of the self-energy
suggest 
a tendency of equalization between the layers for parameter set A, but not
for B.

Nevertheless, the two planes turn out to be  disconnected
for  regions far from the Fermi surface, e.g.  close to $\vec k = (0,0)$,
 even for data set A.
\begin{figure}
\centerline{\epsfig{file=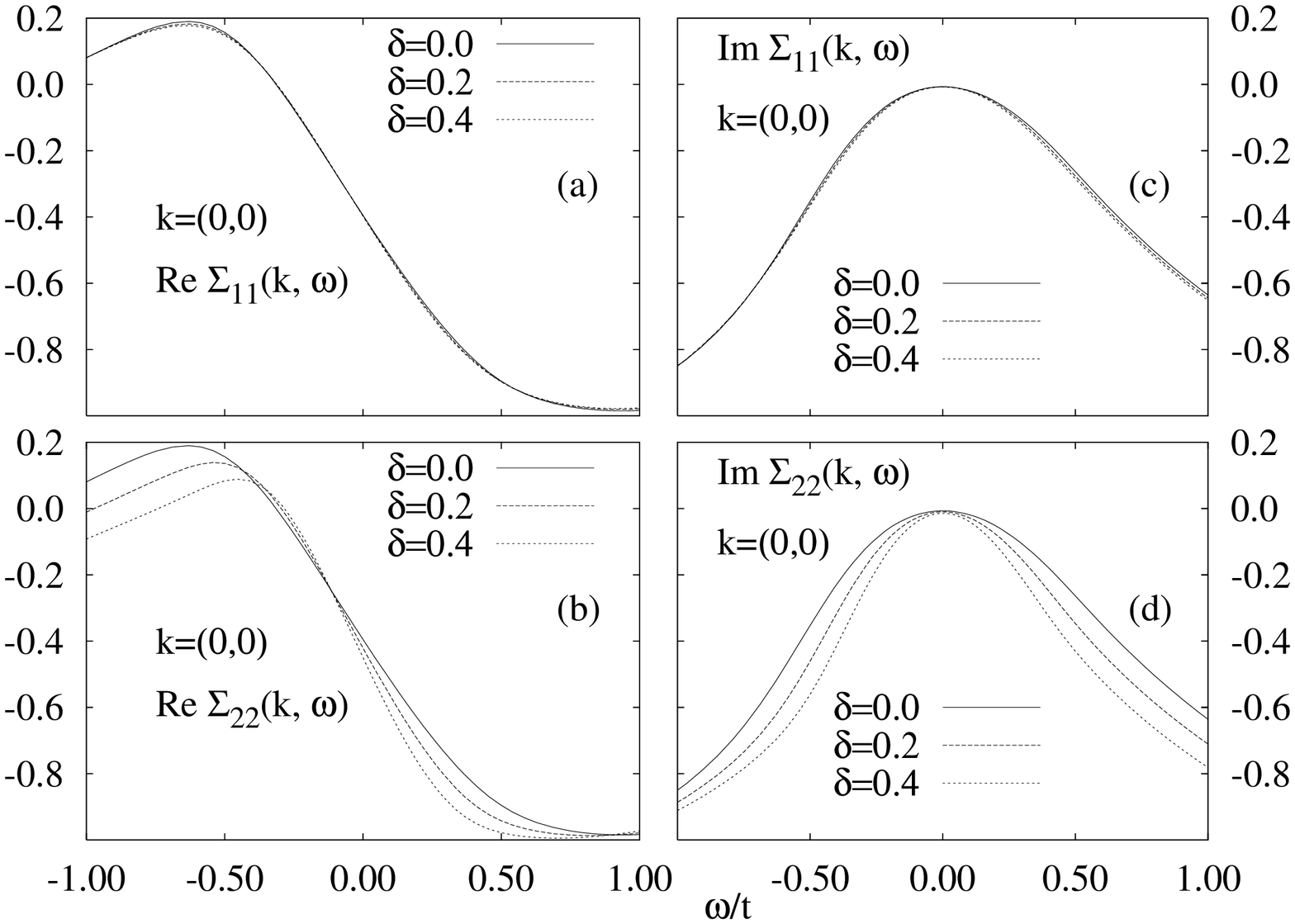,width=9cm}}
\caption{\label{fig:tdl_selfenergy_00}
Real and imaginary part of the two diagonal 
elements of the self-energy for parameter set A and for $\vec k=(0,0).$}
\end{figure}
The real and imaginary parts of the self-energy 
for parameter set A and
 $\vec k=(0,0)$ are
presented in \rfig{fig:tdl_selfenergy_00}.  As can be seen in part 
(a) and (c) of this figure,  changes in the on-site energy $\delta$ of the
second plane   have almost
no effect on the first plane.
The two planes are thus  connected only for momenta 
close to the Fermi surface
which are strongly affected 
by
antiferromagnetic fluctuations at low energy and in particular for
 {\em hot} quasiparticles states around
$(\pi,0)$.

\begin{figure}
\centerline{\epsfig{file=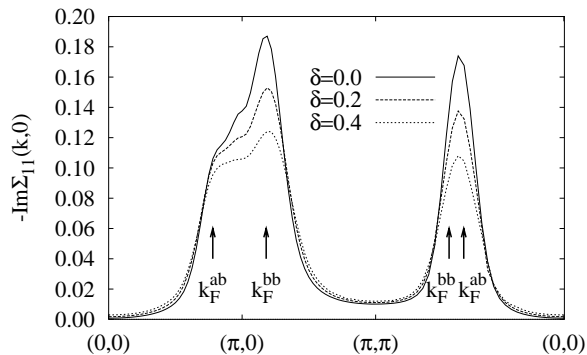,width=8cm}}
\caption{\label{fig:tdl_ImS11_path} 
Scattering rate in the first plane $-{\rm Im}\Sigma_{11}(\vec k, \omega=0)$
along the standard path in the Brillouin zone 
for parameter set A with  $x_1=0.08$,
$T=0.02t$, $\delta/t=0.0, 0.2, 0.4$. The arrows indicate crossing of the anti-bonding $(\vec
k_F^{ab})$ and bonding $(\vec k_F^{bb})$  Fermi surfaces (see also \rfig{fig:fermiflaechen}).}
\end{figure}
To further elucidate the momentum-resolved equalization effects for parameter
set A, 
we show in \rfig{fig:tdl_ImS11_path} the scattering rates at the Fermi energy for the first
plane, namely $-{\rm Im}\Sigma_{11}(\vec k,\omega=0)$ along the standard path
in the Brillouin zone,
for different values of the on-site energy
$\delta$ of the second plane. This figure demonstrates that the scattering
rates are strongly modified at the Fermi surface  with strong
effects in the regions close to $(\pi,0)$. On these regions, the large
number of states associated with  the flat bands produces strong
scattering processes, whenever the interaction 
connects
the van Hove regions of the bonding and antibondig band.
We stress 
again that the variation of the first plane is solely caused by its
 correlation with the second plane 
since the doping of the first plane is kept constant.
 The corresponding real part of the self-energy (not shown) 
 also reveals coupling effects for all $\vec k$ close to $ \vec k_F$.

\begin{figure}
\epsfig{file=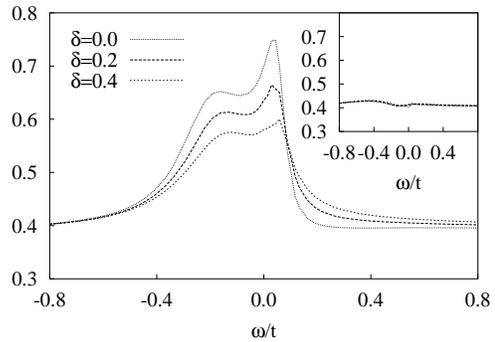,width=7cm}\\
\caption{\label{fig:tperp_eff_vs_ose}
Effective interplane hopping
$t^{\rm eff}_\perp(\vec k, \omega)$ as a 
function of the onsite energy $\delta$ for $t_\perp=0.4t,x_1=0.08,T=0.02t$ for 
parameter set A. The inset shows the results for parameter set B on the
same scale.}
\end{figure}
\begin{figure}
\epsfig{file=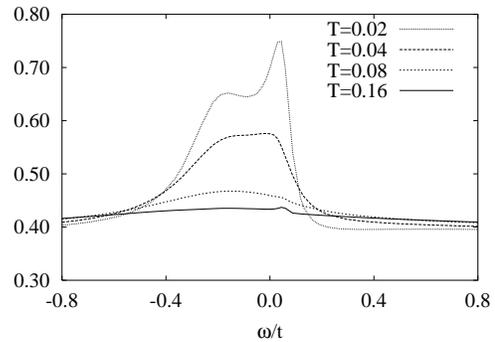,width=7cm}\\
\caption{\label{fig:tperp_eff_vs_T}
Effective interplane hopping 
$t^{\rm eff}_\perp(\vec k, \omega)$ as a function of temperature for
parameter set A with
$x_1=0.08$, $\delta=0.0$. Note the stronger dependence on $T$ than on
$\delta$ when compared to \rfig{fig:tperp_eff_vs_ose}.}
\end{figure}
We now address the question of how much the  hopping 
$t_\perp$
between the planes
 is affected by the interaction.
Dyson's equation (\req{eq:tdl_dyson}) suggests a momentum- and 
energy-dependent  effective interlayer hopping according to
\be
t^{\rm eff}_\perp(\vec k, \omega) = t_\perp + {\rm Re}
\Sigma_{12}(\vec k,\omega) \;.
\ee
Like the real and imaginary parts of the diagonal elements of $\Sigma(\vec
k, \omega)$, this quantity is strongly affected for $\vec k \approx
(\pi,0)$, therefore, we restrict ourselves to  this momentum.

In \rfig{fig:tperp_eff_vs_ose} we show $t^{\rm eff}_\perp(\vec k, \omega)$ 
for parameter set A at a
fixed temperature $T=0.02t$ for a series of on-site energies $\delta$,
while the inset presents the same quantities  for  parameter set B.
Looking at \rfig{fig:tperp_eff_vs_ose}, we observe that $t_\perp$ is
strongly
renormalized close to the chemical potential $(\omega=0)$, and even
more important, that the  antiferromagnetic fluctuations 
again enhance the interlayer connection, in this case represented by
the effective hopping $t^{\rm eff}$.
On the contrary, $t_\perp$ is essentially unrenormalized, 
independently of the value of $\delta$, in the presence of the
 weak antiferromagnetic fluctuations in
parameter set B (see inset of the figure). 
This result seems to stress once more the fact that the planes are strongly
connected for the ``antiferromagnetic'' parameter set A, whereas they
are essentially independent for the parameter set B.
We thus focus our attention on parameter set A where the effects are stronger.
For $\delta=0$ the effective
hopping $t^{\rm eff}_\perp((\pi,0),0) \approx 0.8t$
is roughly twice as large as the bare $t_\perp=0.4t$.
Thus, the hopping between the planes is  {\it amplified} rather than
blocked by 
 electronic correlations. 

\rfig{fig:tperp_eff_vs_ose} 
also shows that the renormalization of $t_\perp$  
decreases with  increasing on-site energy $\delta$, which causes the
second plane to be less magnetic. However, an increase of $\delta$ 
has a surprisingly weak effect on $t^{\rm eff}_\perp((\pi,0),\omega)$ when
compared with an increase of the temperature $T$, which is shown in
\rfig{fig:tperp_eff_vs_T}. This strong temperature dependence of $t^{\rm
eff}_\perp((\pi,0),\omega)$ shown in the figure suggests that the
interlayer coupling is related to a small energy scale $\omega^\star$. For high
temperatures, the thermal fluctuations destroy the correlations on the
small energy scale $\omega^\star$.

\begin{figure}
\epsfig{file=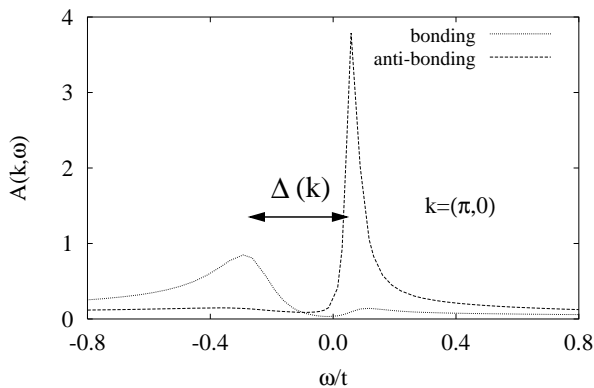,width=8cm}\\
\caption{\label{fig:tdm38_460_A_pi0_band}
Spectral functions $A(\vec k,
\omega)$ of the bonding and anti-bonding bands
at $\vec k=(\pi,0)$  for parameter set A with
$x_1=0.08,\delta=0,T=0.02t$}
\end{figure}
Although the above discussion 
suggests an enhancement of  $t_\perp$ due to interaction effects, the
band splitting between the bonding and the anti-bonding band
$\Delta(\vec k) = \omega^+_{\vec k}-\omega^-_{\vec k}$ goes in the
opposite direction and is reduced with
respect to its bare
value $\Delta_{\rm o}=2t_\perp$, in agreement with previous
conjectures \cite{cl.st.long,mo.el.97,cross.97}.
\rfig{fig:tdm38_460_A_pi0_band} shows the spectral functions $A(\vec
k,\omega)$ of the bonding and anti-bonding band for $\vec k =(\pi,0)$ and
indicates a renormalized band splitting 
$\Delta(\pi,0) \approx 0.4t = 0.5 \Delta_{\rm o}$.  
Thus, while the interlayer hopping seems to be
enhanced by about a factor of 2 at low energies in the presence of
strong antiferromagnetic fluctuations,
 the band
splitting behaves in the opposite way and it 
is  reduced by 
 about the same factor for this parameter set.
This different behavior between $t^{\rm eff}_\perp$ and $\Delta(\vec
k)$ is rather surprising, although it
 may be understood by the following argument.
On the one hand, 
the {\it quasiparticle} interplane hopping {\it without residual
  interaction}, related to 
the off-diagonal energy term
$t^{\rm eff}_\perp$,
is {\it enhanced} due to the fact that quasiparticle of two neighboring 
sites on the two
planes are nearly antiferromagnetically ordered and thus have a larger
amplitude to hop. On the other hand, the {\it whole} hopping
amplitude, related to $\Delta(\vec k)$, is suppressed (by a larger
factor than the enhancement of $t^{\rm eff}_\perp)$, due to the
Hubbard repulsion $U$.

\begin{figure}
\centerline{\epsfig{file=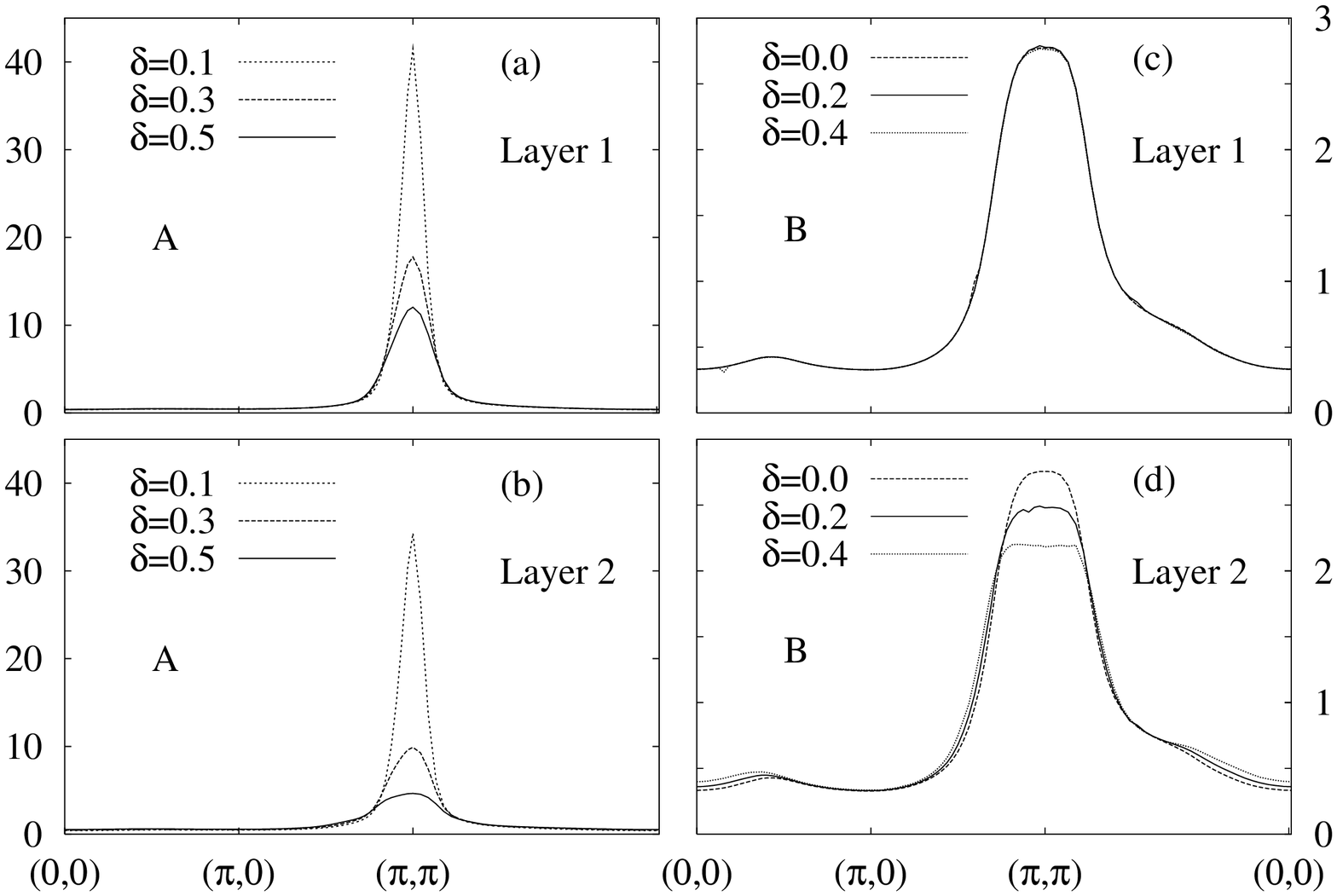,width=9cm}}
\caption{\label{fig:ReXband}%
Static spin susceptibility 
$\chi^{zz}_{ll}(\vec q,\omega=0)$
along the standard path in the Brillouin zone
for the two layers and both parameter sets A and B. ($T=0.02t, x_1=0.08$)}
\end{figure}

Equalization effects between the planes are also observed in
two-particle quantities like the spin response as  
deduced from the spin-spin correlation function. This is  given (in
the layer representation) by
\be
\chi^{zz}(\vec q, \omega)=2\left[1-U \chi(\vec q,
\omega)\right]^{-1} \chi(\vec q, \omega) \; .
\label{eq:tdl_Xzz}
\ee
The static spin-spin correlation function $\chi^{zz}_{ll}(\vec q,
\omega=0)$ along the standard path $(0,0) \to (\pi,0) \to (\pi,\pi) \to
(0,0)$ in the Brillouin zone is
shown in \rfig{fig:ReXband} for $l=1$ and for $l=2$ for 
parameter set A and B.
In the case of parameter set A,
the spin response in both planes is strongly peaked at $\vec
q=(\pi,\pi)$ indicating considerable antiferromagnetism in the Hubbard
planes. However, even more important is the strong dependence of
$\chi^{zz}_{11}(\vec q, 0)$  on $\delta$ which is solely due to the
interplane coupling since the doping in the first plane
is fixed. In agreement with the effects observable in the
single-particle spectrum represented by the self-energy, this clearly
reveals a strong connection between the planes. 
\rfig{fig:ReXband}(a) and (b) also show that the spin response is very
sensitive to a variation of $\delta$ and the antiferromagnetism is
suppressed if $\delta$ is increased.
This is due to the increasing hole concentration in 
the total system which tears it
away from half-filling where  antiferromagnetism is strongest.
The static spin response 
for the data set labeled by B is shown 
in \rfig{fig:ReXband}(c) and (d).
For this choice of parameters, the first plane is again completely disconnected
from the second one since a change of $\delta$ only influences the spin
response of the second
plane while that of the first plane is not  affected at all. 
A clue towards the understanding of the difference between the parameter
sets A and B is already found in the behavior of the spin response.
Comparing \rfig{fig:ReXband} (a)-(b) with \rfig{fig:ReXband} (c)-(d)
reveals a considerable (one order of magnitude) 
smaller value for the static spin response at $(\pi,\pi)$ in the second case.
Furthermore, the spin correlation length $\xi$,
which is the inverse of the half
width at half maximum of $\chi^{zz}(\vec q, 0)$ is much smaller 
(of the order of $1$ lattice spacing)
for
parameter set B. This implies, as expected,
 that the interplane connection is intimately
related to  strong antiferromagnetic correlations in the
planes.

\begin{figure}
\epsfig{file=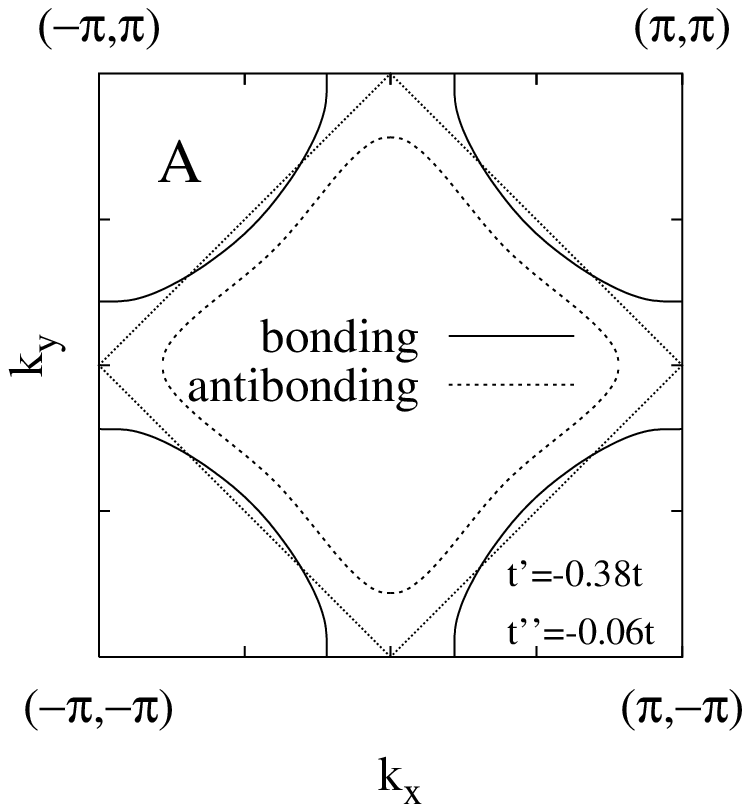,width=4cm}
\epsfig{file=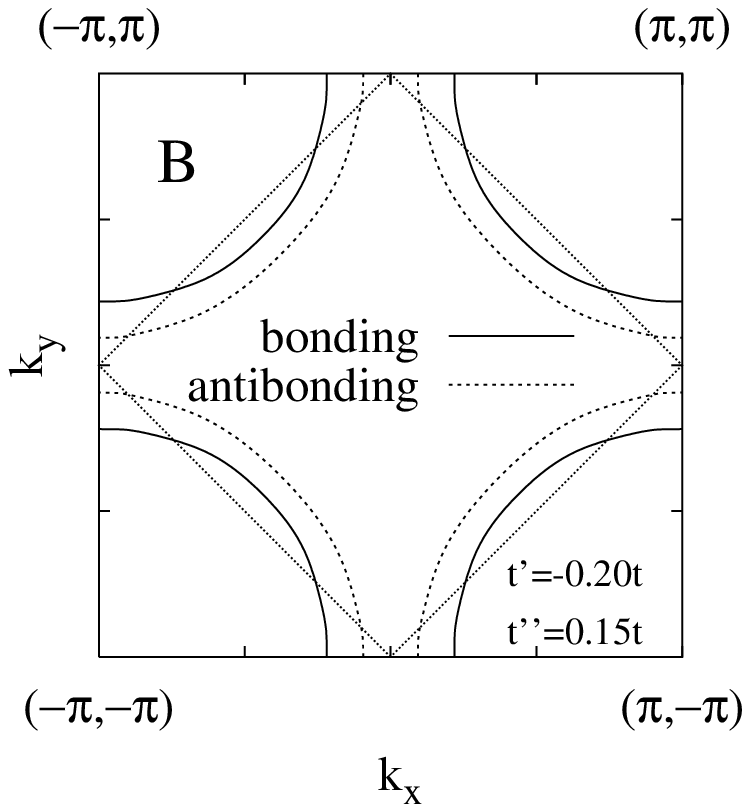,width=4cm}
\caption{\label{fig:fermiflaechen} 
Bonding and anti-bonding Fermi surfaces
for  parameter sets A (left panel) and B (right panel) with
$x_1=0.08$,$\delta=0.2$,$T=0.02t$.}
\end{figure}

The strength of the antiferromagnetic fluctuations are in turn quite
sensitive to the shape of the Fermi surface, especially if large regions 
of the Fermi surface can be linked by the antiferromagnetic momentum
$\vec Q=(\pi,\pi)$ \cite{hi.ar.98.1}.
Keeping this in mind, the differences between parameter set A and B are
caused by their different Fermi surfaces. These are shown in
\rfig{fig:fermiflaechen}. While the 
antibonding and bonding 
Fermi surfaces associated with
parameter set A are 
closed around $(0,0)$ and 
$(\pi,\pi)$, respectively, both Fermi surfaces of 
B  are closed around 
$(\pi,\pi)$.
Even more important is the fact that large regions of
both Fermi surfaces in case A may be connected by $\vec Q$. 
These regions are close to $(\pi,0)$, i.e. close to the van Hove
singularities in the density of states, thus opening various channels for
antiferromagnetic scattering processes.

\section{Relation to experiments}
\label{rela}
We now turn to the question how our theoretical calculations compare with
experimental results.
We thus concentrate our 
attention to NMR experiments on  \ybcoAB\ performed by Stern \ea
\cite{st.ma.94,st.ma.95.1}.
Here, 
 the experimentally relevant quantities 
are the spin-lattice
relaxation time $T_1$ and the Gaussian contribution $T_{2G}$ to the nuclear
spin-spin relaxation time $T_2$.

As pointed out by Shastry\cite{shas.89}
and Mila and Rice\cite{mi.ri.89},
the spin-lattice relaxation time $T_1$ is related to the
spin susceptibility $\chi^{zz}$,  via the expression:
\be
\frac{1}{T_1T}= \lim_{\omega \to 0} \frac{1}{2N} \frac{k_B}{\hbar}
\sum_{\vec q} F_c(\vec q) \frac{{\rm Im}\chi^{zz}(\vec q, \omega)}{\hbar
\omega} \;,
\label{eq:tdl_T1T}
\ee
where $F_c(\vec q)$ is the form factor resulting from the Fourier
transform of the hyperfine interaction
\be
F_c(\vec q)=\left\{A_{ab} + 2 B [\cos q_x + \cos q_y]\right\}^2
\; .
\label{eq:tdl_formfactor}
\ee
Thus, $T_1$ probes the slope of the imaginary part of $\chi^{zz}(\vec q,
\omega)$ for $\omega \to 0$.
In contrast to this, the Gaussian component of the transverse relaxation time 
$T_{2G}$ depends on the static susceptibility 
and is  given by
\bea
T_{2G}^{-2}&=&\frac{0.69}{128 \hbar^2}
\Bigg[ \frac{1}{N} \sum_{\vec q} F_{\rm eff}^2(\vec q)
\chi^{zz 2}(\vec q,0) \nonumber \\
&&-\bigg(\frac {1}{N}
\sum_{\vec q} F_{\rm eff}(\vec q)\chi^{zz}(\vec q,0)\bigg)^2\Bigg]
\label{eq:tdl_T2G}
\eea
as pointed out independently by Takigawa\cite{taki.94} and Thelen and
Pines\cite{th.pi.94}.
While \req{eq:tdl_T2G} applies for the diagonal elements of $\chi^{zz}_{ll}$,
i.e. for the in-plane relaxation rates, the corresponding
inter-layer relaxation rate 
$1/T^{12}_{2G}$ needs not to be 
corrected by the self-interacting hyperfine interaction and thus 
reads:\cite{mo.ri.94,mi.mo.96}
\be
\big[T^{12}_{2G}\big]^{-2} =  \frac{0.69}{128 \hbar^2}
\frac{1}{N} \sum_{\vec q} F_{\rm eff}^2(\vec q)
\big[\chi^{zz}_{12}(\vec q,0) \big]^2
\label{eq:tdl_T2perp}
\ee
The form factor $F_{\rm eff}(\vec q)$ in the last two equations is simply
obtained form
Eq.~(\ref{eq:tdl_formfactor}) by replacing $A_{ab}$ with $A_{c}$.
Although we are  studying  inequivalent Hubbard
planes, we assume for simplicity
that the hyperfine constants, which are usually extracted
from Knight shift experiments, are identical in both planes.
 This is supported by the rather moderate variations of the hyperfine coupling
constants for different cuprate superconductors.
Quantitative differences may be
obtained
if one takes into account
different hyperfine constant, although the temperature dependence
should not change.
To be specific,
we adopt here the values recently given in an analysis of NMR experiments
on YBCO and LSCO by Barzykin and
Pines~\cite{ba.pi.95} and set
$A_{ab}=0.84 B, A_c=-4B$ and  the energy scale
$B=3.82\times 10^{-7} {\rm eV}$.
Note, that since NMR probes the local environment of the spins, all
momenta $\vec q$ contribute to the relaxation times although the main
contributions come from the regions $\vec q \cong (\pi,\pi)$.
Hence, the behavior of the NMR relaxation times are strongly
influenced  by the
antiferromagnetic response of the system.

\begin{figure}
\epsfig{file=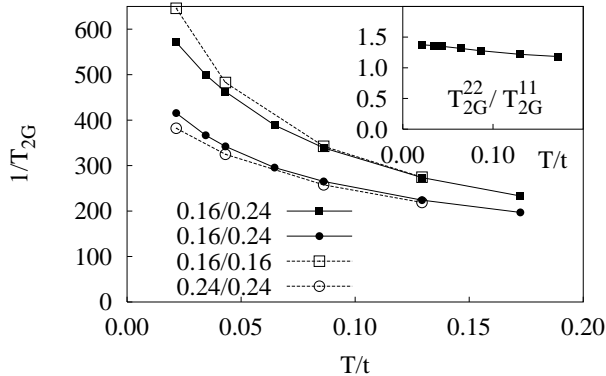,width=8cm}\\
\caption{\label{fig:tdl_T2G}
Temperature dependence of the
Gaussian component of the spin-spin relaxation
rate $T_{2G}$ of a bilayer system consisting of two different layers
with doping
 $x_1=0.16$ and $x_2=0.24$, respectively
$(\delta=0.4)$.
For comparison, we show also the result
for two bilayer systems with equivalent layers with doping
$x_1=x_2=0.16$ and $x_1=x_2=0.24$, respectively. The other parameters
for all curves are:
$t'=-0.38t$,$t''=-0.06t$. The inset shows the ratio
$T^{22}_{2G}/T^{11}_{2G}$ for the case of inequivalent layers.}
\end{figure}

We start the discussion with the transverse relaxation time $T_{2G}$.
In \rfig{fig:tdl_T2G} we show $1/T_{2G}$ as a function of temperature for
both planes of the 
 system with {\it inequivalent} layers
in comparison with the corresponding data for
two corresponding
bilayer systems with {\it equivalent} layers, one with the same doping
as the first layers and one with the same doping as the second layer
of the first system.
 These results for the different layers 
are obtained with \req{eq:tdl_T2G}
by substituting $\chi^{zz}$  with element (11) or (22) from 
\req{eq:tdl_Xzz}.
The filled (open) symbols in this figure are related to the system with
inequivalent (equivalent) layers. Furthermore, the squares represent the
data for planes with a hole doping of $x=0.16$ and the bullets 
planes with doping  $x=0.24$. 
From \req{eq:tdl_T2G} it is seen that a system with rather strong magnetism
and hence large values of $\chi^{zz}(\vec q)$
exhibits large relaxation rates $1/T_{2G}$. This explains the differences
between the layers with $x=0.16$ and $x=0.24$, i.e. with different doping
levels, whereby the heavily doped plane shows a smaller relaxation
rate. 
The most striking result is 
 that data for a plane with a given doping also depend on whether
that plane is coupled with an equivalent one  or with a 
more or less doped one. 
The heavily doped plane $(x_2=0.24)$ of the  system with inequivalent layers 
shows stronger magnetic fluctuations than the plane in the corresponding
system with equivalent bilayers $(x_1=x_2=0.24)$. Similarly, the magnetism of the 
lower doped plane $(x_1=0.16)$ is reduced 
with respect to the corresponding equivalent-layer system $(x_1=x_2=0.16)$
due to the coupling to a
stronger doped plane.
Thus, the magnetic fluctuations of the 
two inequivalent  planes with different carrier concentration
tend to be equalized by
 interplane coupling effects.
A related effect has been detected also in Quantum-Monte-Carlo simulations of
coupled Hubbard planes 
carried out
by Scalettar et al. \cite{sc.ca.94}, in which the two planes 
have different chemical potentials and one
plane is adjusted to  half filling.
In this case, the antiferromagnetic susceptibility of  the half-filled plane is
reduced by the coupling with the  doped plane.
The authors explain this by
the fact that  processes in which holes hop from
the  doped layer into the half filled one 
are energetically more favorable than
virtual hopping processes associated with the magnetic exchange
$J\propto t^2/U$.
\begin{figure}
\epsfig{file=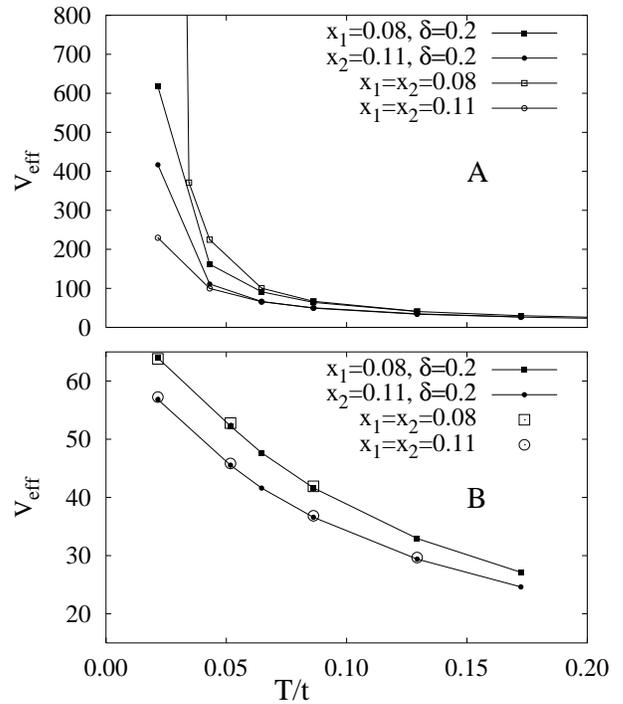,width=8cm}\\
\caption{\label{fig:Veff}%
Effective interaction $V_{\rm eff}(\vec Q,\omega=0)$ for a  system  with 
inequivalent layers ($x_1=0.08$ and $x_2=0.11$) in comparison with the corresponding
 bilayer systems with $x_1=x_2=0.08$ and $x_1=x_2=0.11$. The upper panel shows the results
for parameter set A and the lower for  B.}
\end{figure}
The
effects discussed above 
are also visible in other quantities, like the effective
interaction $V_{\rm eff}(\vec q,\omega)$, which is proportional to the spin
susceptibility in our approximation. For this quantity, we can see
 that the stronger the antiferromagnetic fluctuations are in
the planes, the stronger is the tendency to  equalization of the
two planes.
In  \rfig{fig:Veff}(a) we show  $V_{\rm eff}(\vec q=(\pi,\pi),\omega=0)$
as a function of temperature 
for the parameter set A 
as measured in the two 
layers of the system with inequivalent layers.
In the same figure, we also report for comparison the data for the
two corresponding systems with equivalent layers with their doping
adjusted to the one
of each of 
the two layers of the  first system.
In \rfig{fig:Veff}(b) we show the same comparison for the systems with
parameter set B.
The difference is striking. For parameter set A, $V_{\rm eff}$
 of each of the two layers in the  system with inequivalent layers is
considerably different from $V_{\rm eff}$ in  the
corresponding system
with 
equivalent layers  and tend to be equalized for the two layers.
On the other hand, for the less antiferromagnetic parameter set B, 
$V_{\rm eff}$ calculated on a given layer of the system with
inequivalent layers
 is essentially the same as the one calculated on the 
system with the same doping and equivalent layers.

\begin{figure}
\epsfig{file=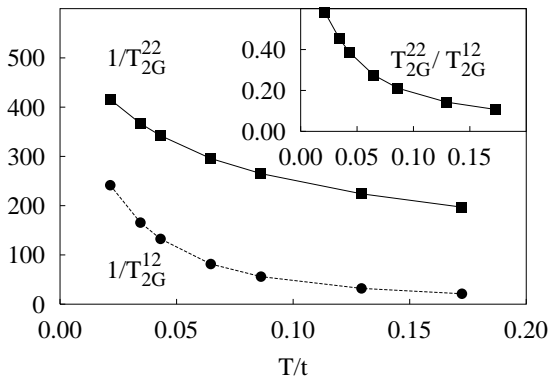,width=8cm}\\
\caption{\label{fig:SEDOR_T2G}
Intra- and inter-plane spin-spin relaxation rates
$1/T_{2G}^{22}$ and $1/T_{2G}^{12}$, respectively.
The inset shows the ratio between the two
relaxation rates $(x_1=0.16, x_2=0.24, \delta=0.4)$.
}
\end{figure}
Qualitatively, the experimental situation for the system with
inequivalent layers,
  \ybcoAB, appears to be described by parameter set A.
Indeed, NMR experiments by Stern et al. \cite{st.ma.95.1} 
on $1/T_{2G}$ for the two layers of
\ybcoAB\ and for the two associated 
systems with equivalent layers,
\ybcoA\ and 
\ybcoB, show the same  behavior as observed in 
\rfig{fig:tdl_T2G} and in Fig. \rfig{fig:Veff}(a), if one identifies the
behavior of $1/T_{2G}$ with the one of $V_{\rm eff}$.
Both experimental and theoretical results show a strong
increase of $1/T_{2G}$ as the temperature is lowered. However, we do not
obtain the decrease of $1/T_{2G}$  below $T_{sg} \approx 100{\rm K}$ which is
attributed to the opening of a gap in the spin excitation spectrum,
since this region 
 is probably unaccessible by our approximation.
Another 
 experimental observation is that the spin-lattice
relaxation rate $1/T_{2G}$  has the same temperature dependence
in the two planes of 
\ybcoAB. 
This has been deduced from the ratio  $R=(1/T^{124}_{2G})/(1/T^{123}_{2G})$ of 
 $1/T_{2G}$ in the two planes,
which turned out to be temperature independent  and approximately
$R\approx$~1.4--1.5.
Since the 124 plane in the coupled layer structure 
of  \ybcoAB\ is the one with lower doping,
 the $\rm CuO_2$-layer from the 123 block corresponds to
the second plane in our theoretical study.
The calculated values for $R=T^{(22)}_{2G}/T^{(11)}_{2G}$
are  presented in the inset of 
\rfig{fig:tdl_T2G}. It turns out that  $R$ 
is almost independent of the temperature $T$, in 
agreement with the experimental finding.
\begin{figure}
\centerline{\epsfig{file=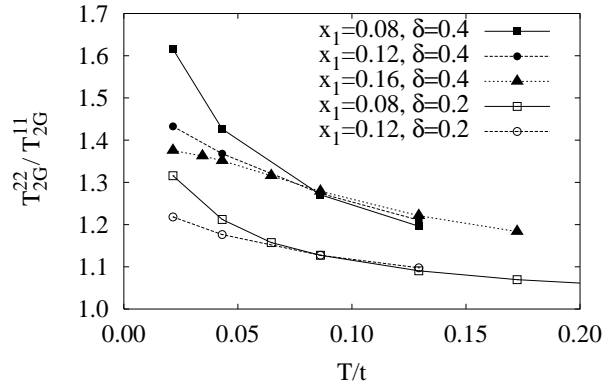,width=8cm}}
\caption{\label{fig:T2G_multi_ratios}%
The ratio $R(T)=T^{(22)}_{2G}/T^{(11)}_{2G}$
for parameter set A and various values of $x_1$ and $\delta$.}
\end{figure}
In fact, a systematic study 
shows that the ratio $R$ is
mainly controlled by the on-site energy $\delta$ and thus by the difference
of particle densities, $\delta n=n_1-n_2$, between the two planes. On the
other hand, the temperature dependence of $R(T)$ is sensitively
related to
the doping of the first layer: for low doping  $x_1$, $R(T)$ shows an
upturn with decreasing $T$ while it is essentially constant for large
values of $x_1$ (see \rfig{fig:T2G_multi_ratios}). 
Thus, the experimental data for $R(T)$ 
imply a rather large doping $x_1$ of
the first layer  in connection with a large on-site energy difference
$\delta$ or, equivalently, filling difference $\delta n$.
On the other hand, the doping  cannot be too large because otherwise
no tendency to equalization  would be observable. An optimal choice for the doping
levels in the coupled system turns out to be $x_1=0.16$ and
$x_2=0.24$, as shown in \rfig{fig:tdl_T2G}.

For the same parameter choice, we  compare the temperature dependence of
the in-plane relaxation rate $1/T_{2G}^{22}$ with the inter-plane 
one $1/T_{2G}^{12}$,
which is calculated using \req{eq:tdl_T2perp}. The latter quantity has
been measured
using NQR-SEDOR experiments \cite{st.ma.95.2}, as suggested by Monien
and Rice\cite{mo.ri.94}. 
 The apparent feature in
these experiments is
that the inter-plane relaxation rate  increases faster for decreasing temperature
than the in-plane one, as seen from the temperature dependence of 
the ratio 
$R_{\rm SEDOR}(T)=T_{2G}^{22}/T_{2G}^{12}$.
Our theoretical calculations displayed in
 \rfig{fig:SEDOR_T2G}  clearly
 reproduce the qualitative behavior observed 
experimentally.
 However, we do not observe the saturation effect
for very low temperatures
as seen in experiments and which is most probably due to the opening
of the spin gap.

\begin{figure}
\epsfig{file=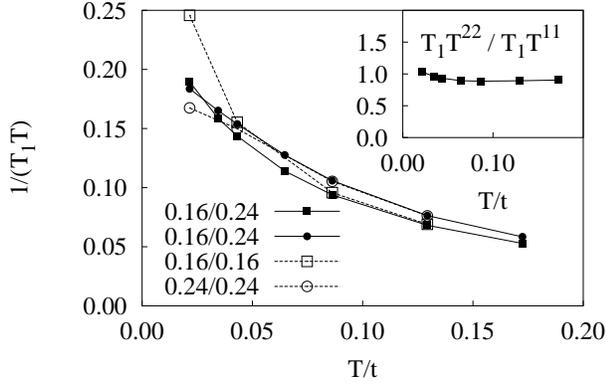,width=8cm}\\
\caption{\label{fig:tdl_T1T}
Temperature dependence of the spin-lattice
relaxation rate $1/(T_1T)$ of a system with inequivalent layers compared to
the corresponding systems with equivalent layers and the same doping
concentration, the inset  shows the ratio of 
$1/(T_1T)$
 for the
two layers of the
 inequivalent-layer system (cf. \rfig{fig:tdl_T2G}).}
\end{figure}

A similar study of the spin-lattice relaxation time $T_1$ on Cu-sites
shows less clear
coupling effects between the two layers, both theoretically and
experimentally.
The experimental results \cite{st.ma.94} show almost no difference between
the $1/(T_1T)$ data for the 123/124 layers in \ybcoAB\ and 
the corresponding layers in the \ybcoA\ and \ybcoB\ systems, respectively.
Deviations can be seen only for rather low temperatures $T\lesssim T_{sg}$.
Above $T_{sg}$, the ratio $R_1(T)=(T_1T^{123})/(T_1T^{124})$ 
deduced from the experiments is again essentially
temperature independent and is about $1.4$.
In \rfig{fig:tdl_T1T} we show the  calculated results for
$1/(T_1T)$ obtained with \req{eq:tdl_T1T} for the same parameters as in
\rfig{fig:tdl_T2G}. Here, we observe that the coupling effects between
the planes are not as strong as for the spin-spin relaxation rate
$1/T_{2G}$.
 In fact, especially for the plane with the  larger doping
 ($x=0.24$) 
$1/(T_1T)$ seems to deviate from the  value 
of the system with equivalent layers
 only for
rather small temperatures $T/t \approx 0.05$. 
On the other hand, the plane with  lower doping
($x=0.16$) shows deviations already for higher temperatures
$T/t\approx 0.08$. 
Even more surprising is the  result for the ratio $R_1(T)$,
shown in \rfig{eq:tdl_T1T}. Here, the plane with the higher doping
shows a larger spin-lattice relaxation rate $1/(T_1T)$, in
contradiction with the simple expectation that lower doping should result in
stronger antiferromagnetic fluctuations.
 The expected value $R_1>1$ is  restored
only for low $T\lesssim 0.04 t$ in our calculation.
The deviation from the experimental ratio
$R_1\approx 1.4$ 
may be caused by the
incorrect assumption that the hyperfine interaction constants are the same
in both planes.

\begin{figure}
\centerline{\epsfig{file=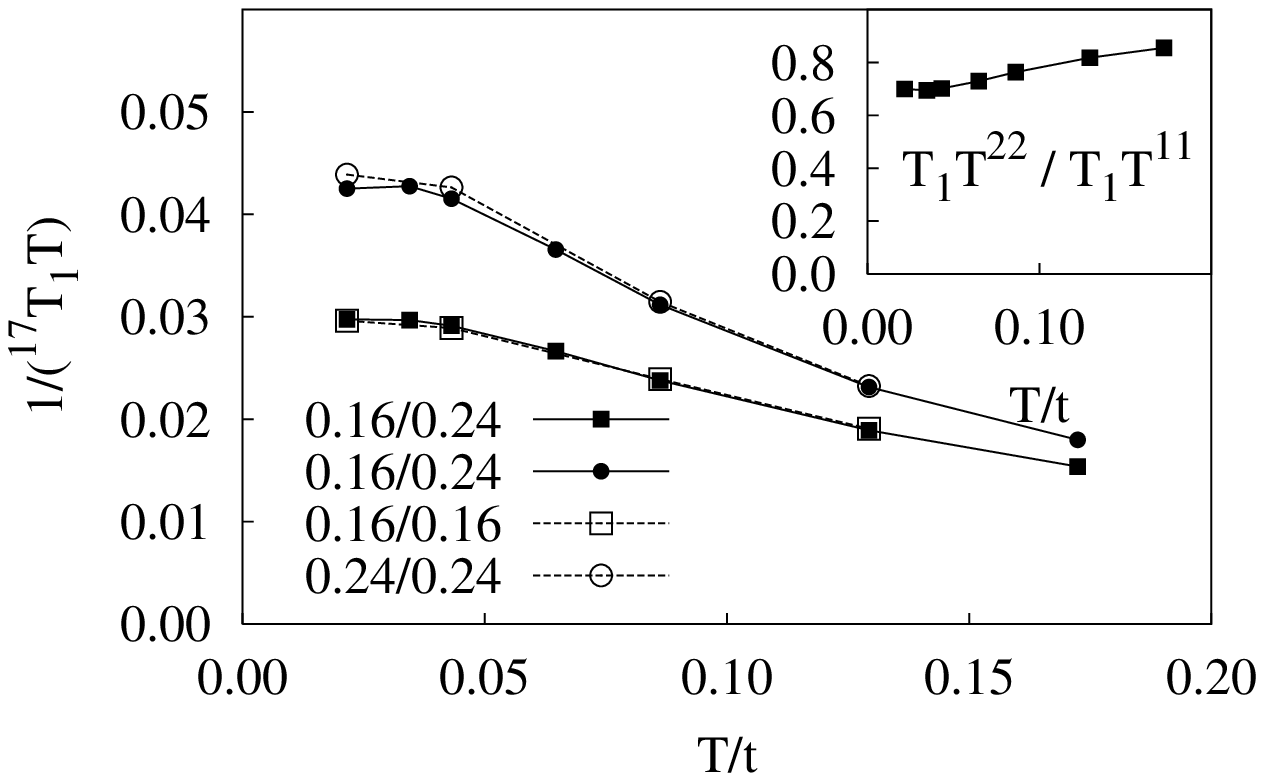,width=8cm}}
\centerline{\epsfig{file=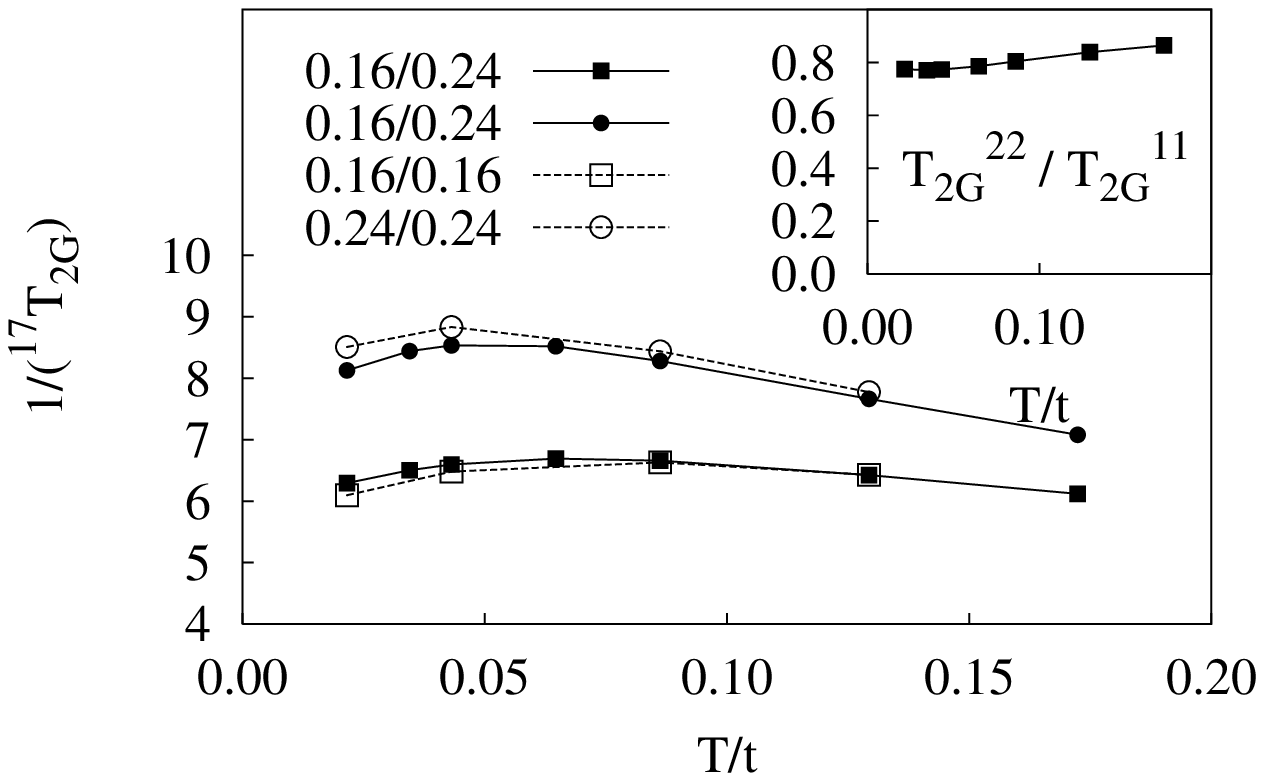,width=8cm}}
\caption{\label{fig:tdl_T1TOX}
Temperature dependence of the spin-lattice
relaxation rates $1/(^{17}T_1T)$ 
and of the spin-spin relaxation rates $1/(^{17}T_{2G})$ 
on oxygen sites. Like in
\rfig{fig:tdl_T2G} and \rfig{fig:tdl_T1T}, the results for a system with
inequivalent layers are compared with the corresponding  systems with
equivalent layers and the same doping  (the parameters are the same as
in Fig. \ref{fig:SEDOR_T2G}).}
\end{figure}
Finally, we have calculated the  relaxation times
$^{17}T_1$ and $^{17}T_{2G}$ 
on
oxygen sites. Like the corresponding relaxation times on copper sites, these
quantities are related to the slope for $\omega \to 0$
of the imaginary part of the spin susceptibility, and by its real
part, respectively. The appropriate form factor 
which enters Eqs.\ref{eq:tdl_T1T},\ref{eq:tdl_T2G}  and which is given by 
$^{17}F(\vec q)=2 C^2 [1+0.5 [\cos(q_x)+\cos(q_y)] ]$
with two different constants for 
$T_1T$ and $T_{2G}$,
 suppresses scattering events  transfering
 momentum $\vec q=(\pi,\pi)$ 
related to magnetic fluctuations
and instead favors those with $\vec q$ around
$(0,0)$. The spin-lattice and spin-spin relaxation times  on oxygen sites
hence mainly probe the center of the Brillouin zone.
We already know from the static spin susceptibility shown in
 \rfig{fig:ReXband}
that the equalization effects
are much weaker for $\vec q \cong (0,0)$ as compared with  $(\pi,\pi)$. 
This weaker connection between the layers at $\vec q \cong (0,0)$ 
is equally pronounced for the dynamical susceptibility and 
thus  leads to  
the oxygen  relaxation rates
$1/(^{17}T_1T)$  and
$1/(^{17}T_{2G})$
shown in \rfig{fig:tdl_T1TOX}.
 Here, we find that the layers of the system consisting of two inequivalent
layers behave almost like the corresponding planes in the systems with
equivalent layers and the two inequivalent layers are essentially 
disconnected. 
 We thus predict  that, within a purely magnetic scenario,
 experimental measurements of 
 the  oxygen
relaxation times in
123 and 124-layers of \ybcoAB\
should behave, as a function of temperature,
like those of the corresponding layers in \ybcoA\ and \ybcoB, respectively.
In other words, if one would detect a different behavior of 
$1/(^{17}T_1T)$ in 
\ybcoAB\
compared to the two parent compounds. it would be a strong indication for  a
nonmagnetic coupling of the two layers.

\section{Conclusions}
\label{conc}
In summary, we have studied a microscopic model consisting of two
inequivalent Hubbard planes which are connected by an interlayer hopping
$t_\perp$. We have shown that magnetic and single-particle fluctuations of
the two layers are connected and tend to be
equalized if the antiferromagnetic fluctuations within the layers are
strong. 
If the antiferromagnetic correlation length is less than 1--2 lattice spacings,
which happens for high temperatures, large doping or bandstructure
parameters with inefficient magnetic coupling between the two
Fermi surface sheeds, we find that the two inequivalent layers are disconnected
and keep their individual properties. However, once the antiferromagnetic correlations in the layer
with smaller charge carrier concentration is sufficiently large, the 
single-particle excitations
for momenta close to the Fermi surface and, in particular,
around the hot-spots as well as the magnetic excitations of
the two layers are strongly connected.
The whole system reacts, despite its inhomogenious charge density, magnetically
as a single system.
In this case, the interlayer antiferromagnetic 
susceptibility, measured by $T_{2G}^{12}$, increases for decreasing 
temperatures more strongly than the individual inplane susceptibilities.
These trends are in agreement with the experimenmtal 
observation by Stern {\em et al}~\cite{st.ma.95.2}
demonstrating that it is sufficient, for an understanding of the  magnetic 
interlayer coupling, to use a  single-particle interlayer hopping element.
Furthermore, we expect from our analysis of the oxygen NMR relaxation rate
that, in distinction to the Cu-relaxation rates,  the  
magnetic connection between the two
layers  will barely be visible, even for strongly underdoped systems.
This phenomenon could be used to separate the small contributions due
to antiferromagnetic correlations from the dominant, rather conventional, contribution 
of the ${\bf q} \approx 0$ dynamical spin susceptibility.
Finally, in a futher step, based on the findings of this paper,
it is of interest to investigate the 
behavior of two coupled, but inequivalent layers in the
superconducting state (some work in this direction, although at much
higher temperatures, has been carried out in Ref. \onlinecite{sc.ca.94}).
This may reveal, why the  material \ybcoAB\  exhibits a higher T$_c$ than 
both of the corresponding parent compounds it consists of.

This work has been supported in part by the Science and Technology
Center for Superconductivity through NSF-grant DMR91-20000, 
   the Deutsche Forschungsgemeinschaft (J.S.), the 
EC-TMR
  program  ERBFMBICT950048 (E.A.), and by FORSUPRA (G.H. and W.H.).
It is our pleasure to thank  H. Monien, D. Pines, R. Stern, M. G. Zacher, and R. Eder
for  helpful discussions. The calculations were performed at the ZAM in
J\"ulich and  the LRZ in Munich.

\ifx\undefined\andword \def\andword{and} \fi 
\ifx\undefined\submitted \def\submitted{submitted} \fi 
\ifx\undefined\inpress \def\inpress{in press} \fi 
\def\nonformale#1{#1}
\def\formale#1{}
\def\spa{} \def\spb{}
\def\spa{\ifpreprintsty\else\vspace*{-.5cm}\fi} 
\def\spb{\ifpreprintsty\else\vspace*{-1.6cm}\fi} 
\spa

\end{multicols}

\end{document}